\documentclass[12pt]{elsart} 
\usepackage{psfig,epsfig}
\usepackage[usenames]{color}
\journal{Nuclear Physics B}   
 
\def\np#1#2#3{    {\it Nucl. Phys. }{\bf #1} (#2) #3} 
 
\def\pl#1#2#3{    {\it Phys. Lett. }{\bf #1} (#2) #3} 
 
\def\pr#1#2#3{    {\it Phys. Rev. }{\bf #1} (#2) #3}

\def\zp#1#2#3{    {\it Zeit. f\"ur Physik }{\bf #1} (#2) #3}

\def\re#1{{\mathrm Re}\left\{#1\right\} }
\def\im#1{{\mathrm Im}\left\{#1\right\} }  
\newcommand{\com}[1]{ \par }

\def\evg{\, g_{eV}^\gamma} 
\def\tvg{\, g_{\tau V}^\gamma} 
\def\evz{\, g_{eV}^Z} 
\def\eaz{\, g_{eA}^Z} 
\def\tvz{\, g_{\tau V}^Z} 
\def\taz{\, g_{\tau A}^Z} 
\def\tz{\, g_{\tau T}^Z} 
\def\tg{\, g_{\tau T}^\gamma} 
\def\evg2{(g_{eV}^\gamma)^2} 
\def\tvg2{(g_{\tau V}^\gamma)^2} 
\def\evz2{(g_{eV}^Z)^2} 
\def\eaz2{(g_{eA}^Z)^2} 
\def\tvz2{(g_{\tau V}^Z)^2} 
\def\taz2{(g_{\tau A}^Z)^2} 
\def\tz2{(g_{\tau T}^Z)^2} 
\def\tg2{(g_{\tau T}^\gamma)^2}

\def\re#1{{\mathrm Re}\left\{#1\right\} } 
\def\im#1{{\mathrm Im}\left\{#1\right\} }

\newcommand{\beq}{\begin{equation}} 
\newcommand{\eeq}{\end{equation}} 
\newcommand{\bi}{\begin{itemize}} 
\newcommand{\ei}{\end{itemize}} 
\newcommand{\bea}{\begin{eqnarray}} 
\newcommand{\eea}{\end{eqnarray}} 
\newcommand{\bes}{\begin{eqnarray*}} 
\newcommand{\ees}{\end{eqnarray*}}

\begin{document} 
\begin{flushright}FTUV-04-0421\end{flushright}
\begin{frontmatter} 
\title{CP violation and electric-dipole-moment at low energy  $\tau$-pair production} 
\author[Valencia]{J. Bernab\'eu}, 
\author[Montevideo]{G. A. Gonz\'alez-Sprinberg} and  
\author[Valencia]{J. Vidal} 
\address[Valencia]{Departament de F\'{\i}sica Te\`orica  
Universitat de Val\`encia, E-46100 Burjassot,Val\`encia, Spain\\ 
and\\
IFIC, Centre Mixt Universitat de Val\`encia-CSIC, Val\`encia, Spain} 
\address[Montevideo]{Instituto de F\'{\i}sica,
 Facultad de Ciencias, Universidad de la Rep\'ublica,  Igu\'a 4225, 11400 Montevideo, Uruguay} 
 \begin{abstract} 
CP violation at low energy is investigated at the 
$\tau$ electromagnetic  vertex. High statistics at B factories, 
and on top of the $\Upsilon$ resonances,
 allows a detailed investigation of $CP$-odd 
observables related to the $\tau$-pair production.
  The contribution of the tau electric dipole moment
 is considered in detail.
We perform an analysis independent  from the high energy data by
 means of correlation and 
linear spin observables at low energy.  We show that different 
$CP$-odd asymmetries,
associated to the normal-transverse and 
normal-longitudinal correlation terms 
can be measured at low energy accelerators,
both at resonant and non resonant energies. 
These observables allow to put stringent and 
independent bounds to the tau electric dipole moment
 that are competitive with other high or low energy results. 
\end{abstract} 
\end{frontmatter} 
 
\section{Introduction} 
The time reversal odd electric dipole moment (EDM) of  
the $\tau$ is the source of 
CP violation in the $\tau$-pair production vertex. In the framework of 
local quantum field theories the CPT theorem states that CP violation 
is equivalent to T violation. 
While the electric dipole moments (EDM) of the electron 
and muon have been
extensively investigated both in experiment and theory,
 the case of the tau is somewhat different.  
The tau lepton has a relatively high mass: this means that tau lepton 
physics is expected   to be more sensitive 
to contributions to chirality-flip terms coming from high energy scales and 
new physics. Furthermore, the tau  decays into hadrons, so different 
techniques to those for the (stable) 
electron or muon case are needed in order to measure the dipole moments. 
 There are very precise  bounds on the EDM magnitude 
 of nucleons and leptons, and the most precise one is the 
 electron 
EDM, $d^e_\gamma = (0.07\pm0.07) \times 10^{-26}$ {\it e cm},
 while the looser one is the  $\tau$ EDM \cite{pdg}, 
$-0.22 \, \it{ e \, cm } < Re (d^\tau_\gamma) \times 10^{16} < 0.45\, 
\it{e\, cm}$.
The dipole moments flip chirality and are therefore related to the mass mechanism
of the theory.  
>From the theoretical point of view the 
CP violation mechanisms in many models provide 
a kind of accidental protection in order 
to generate an EDM for quarks and leptons.
This is the case in the CKM mechanism, where EDM and weak-EDM 
are generated only at very high order in the coupling 
constant. This opens a way to  efficiently 
test many models: $CP$-odd observables related to EDM 
would give no appreciable effect from the standard 
model and any experimental signal should be 
identified with physics beyond the standard model.
Following the ideas of \cite{nt} and \cite{heidel}, the tau 
weak-EDM has been studied in $CP$-odd observables \cite{l3,ao} at high energies 
through terms involving spin linearly and spin-spin correlations. Electric 
dipole moment bounds for the tau, from $CP$-even observables such
as total cross sections or decay widths, have been considered in
\cite{paco,grif,masso}.
While most of the statistics 
for the tau pair production was dominated by high energy physics, 
mainly at LEP, nowadays the situation has evolved. 
High luminosity B factories and their upgrades  at resonant energies 
($\Upsilon$ thresholds) have the largest $\tau$ pair samples.
This calls for a dedicated study of the observables related to CP violation 
 and the EDM of the $\tau$ lepton at low energies. 
In this paper we study different observables 
to the ones used, at high energies, in the past.
 For the tau lepton
we present some of them that may
lead to competitive 
results  with the present bounds in the near future.

The paper is organized as follows: in the next section we present 
the effective Lagrangian description for the  EDM,
in section 3 we discuss the low energy observables, in section 4 we consider the
 resonant production energies, in section 5 we show how to measure the imaginary parts 
and finally we conclude with some comments.

\section{{\LARGE $\tau$} EDM  at Low Energies}

The standard model describes with high accuracy most of the physics found in
present experiments. Nowadays, however, neutrino physics
offers a first clue to physics beyond this low energy model \cite{nu}. 
Deviations from the 
standard model, at low energies, can be
parametrized by an effective Lagrangian built with the standard model
particle spectrum, having as zero order term just the standard model
Lagrangian, and containing higher dimension gauge invariant operators  
suppressed by the scale  of new physics, $\Lambda$ \cite{buch}. The leading non-standard 
effects come from the operators with the lower dimension. For 
CP violation those are  dimension six operators. There are only two 
operators of this type that contribute \cite{arcadi} to the tau EDM and weak-EDM:
\begin{eqnarray}
\label{eq:ob}
\mathcal{O}_B = \frac{g'}{2\Lambda^2} \overline{L_L} \varphi \sigma_{\mu\nu}
 \tau_R B^{\mu\nu} ~,& \hspace*{1cm} \mathcal{ O}_W = \frac{g}{2\Lambda^2} \overline{L_L} \vec{\tau}\varphi
\sigma_{\mu\nu} \tau_R \vec{W}^{\mu\nu}  ~.
\end{eqnarray}
Here $L_L=(\nu_L,\tau_L)$ is the tau leptonic doublet, 
$\varphi$  is the Higgs doublet, $B^{\nu\nu}$ and $\vec{W}^{\mu\nu}$ are the 
U(1)$_Y$ and SU(2)$_L$ field strength tensors, and $g'$ and $g$ are the gauge 
couplings. 
 
Other possible operators that one could imagine
reduce to the above ones of Eq.(\ref{eq:ob})  after 
using the standard model equations of motion. 
In so doing, the couplings will be proportional 
to the tau-lepton Yukawa couplings. 

Thus, we write our effective Lagrangian 
\begin{equation}
\label{eq:leff}
\mathcal{ L}_{eff} = i \alpha_B \mathcal{ O}_B + i \alpha_W \mathcal{ O}_W + \mathrm{h.c.}
\label{eq:interaccio}
\end{equation}
where the couplings
$\alpha_B$ and $\alpha_W$ are real. Note that complex couplings do not break
$CP$ conservation and lead to magnetic dipole moments which are not considered
in this paper where we are mainly interested on $CP$-odd observables.

If these operators come from a low energy expansion of a 
renormalizable theory, in the perturbative
regime one expects that they arise only as quantum corrections 
and therefore their contribution must be suppressed. However, this does 
not need necessarily to be 
the case, therefore  we leave the couplings $\alpha_B$ and
$\alpha_W$ 
as free parameters without any further assumption.

In the spontaneous symmetry breaking regime the neutral scalar gets a vacuum expectation
value 
$<\varphi^0>=u/\sqrt{2}$ with $u=1/\sqrt{\sqrt{2}G_F}=246$~GeV, 
and the interactions in Eq.(\ref{eq:interaccio}) can be written in terms of the gauge 
boson mass eigenstates 
$A^\mu$ and $Z^\mu$. Similar results, but for the 
magnetic moments, are
found in \cite{arcadi} where the notation is the same.
The Lagrangian, written in terms
of the mass eigenstates, is
then
\begin{eqnarray}
\mathcal{ L}_{eff}^{\gamma, Z} &=& 
- i \frac{e}{2 m_\tau} \,F^\tau_\gamma \,\overline{\tau} \sigma_{\mu\nu}
 \gamma^5 \tau F^{\mu\nu} -
 i \frac{e}{2 m_\tau} \,F^\tau_Z \,\overline{\tau} \sigma_{\mu\nu}
\gamma^5 \tau
Z^{\mu\nu}
\label{eq:leff_fin}
\end{eqnarray}
where $F_{\mu\nu}=\partial_\mu A_\nu-\partial_\nu A_\mu$ 
  and 
$Z_{\mu\nu}=\partial_\mu Z_\nu-\partial_\nu Z_\mu$ 
are  the abelian field strength tensors of the photon and
  $Z$ gauge boson, respectively.
 We have not written in Eq.(\ref{eq:leff_fin}) some of the terms coming from 
Eq.(\ref{eq:interaccio}) because they 
do not contribute at leading order to the observables we are interested 
in. These terms are a) the non-abelian
couplings involving more than one gauge boson b) the Lagrangian related
to the $CP$-odd $\nu_\tau - \tau - W^\pm$ couplings. 
As usual, we  define the following dimensionless 
couplings
\begin{eqnarray}
F^\tau_\gamma &=& (\alpha_W - \alpha_B) \frac{u
m_\tau}{\sqrt{2}\Lambda^2}~,\label{f1}\\
F^\tau_Z &=&  (\alpha_B s_W^2 + \alpha_W c_W^2) \frac{u
m_\tau}{\sqrt{2}\Lambda^2} \frac{1}{s_W c_W}\label{f2}
 \label{eq:epsilonw}
\end{eqnarray}
where $s_W =\sin\theta_W$ and $c_W = \cos\theta_W$ are the 
sine and cosine of the weak angle.
In this effective Lagrangian approach the same couplings
that contribute to 
 the electric dipole moment form factor, 
$F^{\mathrm new}(q^2)$ also contribute to the electric dipole moment 
defined at $q^2=0$. Only higher dimension operators contribute to the difference
$F^{\mathrm new}(q^2)-F^{\mathrm new}(0)$ and, if  $ |q^2| \ll \Lambda^2$, as
required for the consistence of the effective Lagrangian approach, their effects
will be suppressed by powers of $q^2/\Lambda^2$. This allows us to make no 
distinction between the electric {\it dipole moment} and the electric {\it form factor}
in this paper so we may take the usual definitions of the electric and 
weak-electric dipole moments in terms of the form factors defined in Eqs.(
\ref{f1},\ref{f2}) as:

\begin{eqnarray}
d^\tau_\gamma = \frac{e}{2 m_\tau} F^\tau_\gamma ,& \hspace{1cm} d^\tau_Z = 
\frac{e}{2 m_\tau} F^\tau_Z .
\end{eqnarray}

\begin{figure}[hbtp]
\begin{center}
\epsfig{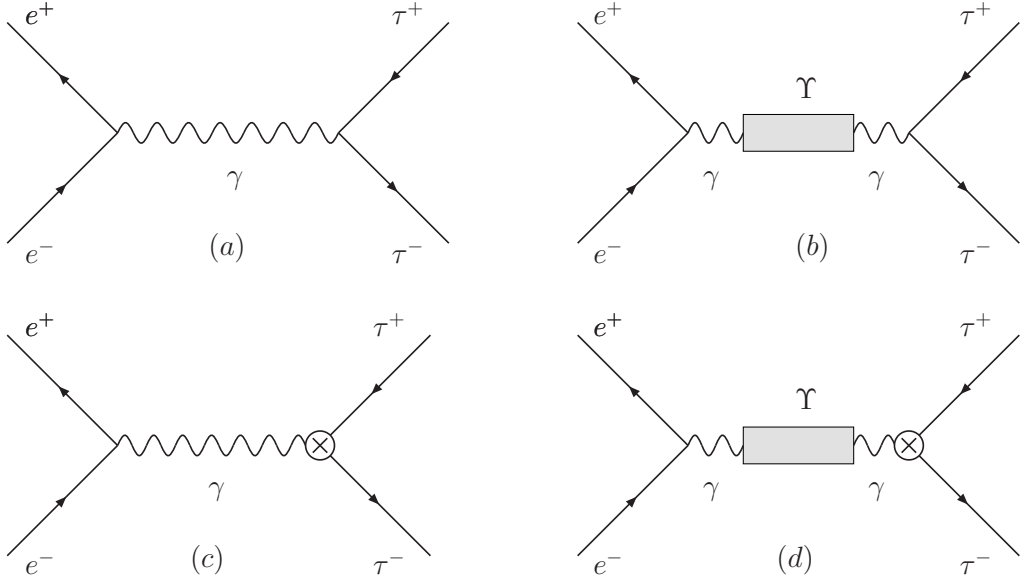}
\end{center}
\caption{Diagrams (a) direct $\gamma$ exchange (b) $\Upsilon$
production (c) EDM in $\gamma$ exchange (d) EDM in $\Upsilon$
production
\label{fig:figura1}}
\end{figure}

The $e^+\, e^- \longrightarrow 
\tau^+ \tau^-$ cross section has contributions coming from
the standard model and the effective Lagrangian 
Eq.(\ref{eq:leff_fin}). At low energies the 
tree level contributions come from
  $\gamma$ exchange (off the $\Upsilon$ peak) or 
$\Upsilon$ (at the $\Upsilon$ peak) exchange in the s-channel. 
The interference with the $Z$-exchange 
($\gamma - Z$,  $\Upsilon - Z$ at the $\Upsilon$ peak) and the 
$Z-Z$ diagrams are suppressed by powers of $\left(q^2/M_Z^2\right)$. The tree level 
contributing diagrams are shown in  Fig.\ref{fig:figura1} where diagrams ($a$) and 
($b$) are standard model contributions, and ($c$)and ($d$) come from beyond the standard
model terms in the Lagrangian. Notice that 
standard model radiative corrections 
that may contribute to $CP$-odd observables (for example, the ones 
                                                 that generate the 
standard model electric dipole moment for the $\tau$) 
come in higher order in the  coupling constant, and at present level of
experimental sensitivity they are not measurable. On these grounds the  bounds on the 
EDM that one may get are just coming from the physics beyond the standard model.

Electric dipole moment effects can be studied at  
leading order in the angular distribution of the 
$e^+e^-  \longrightarrow \tau^+(s_+)\tau^-(s_-)$ differential cross section.
The polarization of the final fermions is determined through the study of the
angular distribution of their decay products. In our analysis we only keep 
linear terms in the EDM, neglecting terms proportional to the mass of the
electron.

When considering the measurement of the polarization of just one of the taus, 
the normal -to the scattering plane- polarization ($P_N$) of each tau 
 is the only
component which is $T$-odd. For $CP$-conserving interactions, the 
$CP$-even term $(s_++s_-)_N$ of the normal polarization only gets 
contribution through the combined 
effect of both an helicity-flip transition and the presence of 
absorptive parts, which are both
suppressed in the standard model. For a $CP$-violating interaction, 
such as an EDM, the $(s_+-s_-)_N$ $CP$-odd term  gets a
non-vanishing value without the need of absorptive parts.

As $P_N$ is even under parity (P) symmetry, the observable sensitive to 
the EDM should also be proportional to a standard axial coupling in 
addition to $d_\gamma^\tau$. This would need a Z-exchange in the s-channel, 
which is suppressed by powers of $\left(q^2/M_Z^2\right)$
at low energies.
In our case, with only $\gamma$ and/or $\Upsilon$ exchange 
(the complete spin density matrix 
can be seen, for example, in the cross section formulas of ref. \cite{500}) 
there
is no such amplitudes and the EDM does not give any contribution to 
the  single normal polarization ($P_N$) of the tau. As a consequence, we 
have to move to other observables associated with spin correlations of 
both taus.

 
Then, the EDM term only shows up in the spin-spin correlation matrix. 
The $T$-odd, $P$-odd Normal-Transverse 
$(\overrightarrow{s}_+\times \overrightarrow{s}_-)_{NT}$ 
and Normal-Longitudinal 
$(\overrightarrow{s}_+\times \overrightarrow{s}_-)_{NL}$ 
spin correlation terms will be  proportional to the EDM interfering with 
photon exchange. These two spin correlations  receive
standard model contributions to their symmetric $CP$-even 
terms 
through absorptive parts generated in
 radiative corrections. At leading order it is the imaginary 
part of the $Z$ propagator that produces a contribution to these correlations
with the interference of the amplitudes of direct 
$\gamma$ and $Z$-exchange. This term is suppressed at low 
energies by the factor  $(\frac{q^2}{M_Z^2} \frac{ \Gamma_Z}{M_Z})$, and
has been calculated in 
\cite{nuria}, so that it can be subtracted if necessary.
 The Transverse-Longitudinal term 
$(\mbox{\boldmath $s$}_+ \times 
\mbox{\boldmath $s$}_-)_{TL}$ is $T$-even, $P$-even and it can 
contain a term proportional to the EDM only through its interference 
with  $Z$ amplitudes
carrying an axial coupling. As  mentioned before, these are suppressed at 
low energies at least by $(q^2/M_Z^2)$, and thereby
not considered in what follows.
 Our aim is to identify genuine $CP$-odd observables linear in the EDM and 
not (additionally) suppressed by either $(q^2/M_Z^2)$ or unitarity 
corrections.

\section{Low energy observables} 

Following the notation of references \cite{arcadi} and \cite{nos}, we 
now show how to  measure the EDM using low energy $CP$-odd observables.  
At low energies and in the  hypothesis stated above, the EDM gives 
contributions to leading order in the Normal-Transverse and 
Normal-Longitudinal correlation terms of the 
$e^+e^-  \longrightarrow \tau^+(s_+)\tau^-(s_-)$ differential cross section.

Working in the center of mass (CM) reference frame we choose the orientation of
our coordinate system so that
the outgoing $\tau^-$ momenta is along the positive $z$ axis and the vector 
$\mbox{\boldmath $p$}_{\tau^-}\times \mbox{\boldmath $p$}_{e^-}$  
 defines  the positive $y$ axis.
The $\mbox{\boldmath $s$}^\pm$  are the $\tau^\pm$  spin vectors in
the $\tau^\pm$ rest system, $s_\pm=(0, s^x_\pm, s^y_\pm, s^z_\pm)$.
This frame has the axes parallel to the CM frame and the 
only difference between them and the CM frame of reference 
is the boost in the $z>0$ direction in the 
case of the $\tau^-$ and the boost in the $z<0$
direction for the $\tau^+$ frame. With this setting, polarization 
along the directions 
$x,y,z$ correspond to what is called transverse (T), normal (N) and longitudinal (L)
polarizations, respectively.

We  consider the $\tau$-pair production in $e^+e^-$ collisions 
though direct $\gamma$ exchange (diagrams (a) and (c) in Fig. \ref{fig:figura1}.). 
In the next section we will show that the basic results of this section still hold
for resonant $\Upsilon$ production.

Let us  assume from now on that the tau
production plane and direction of flight can be fully reconstructed. 
This can be easily done \cite{kuhn} if both $\tau$'s decay semileptonically. Following 
the ideas of \cite{nt,nos} this technique was  applied by the L3-Collaboration \cite{l3} in
the search of bounds on the tau weak electric and magnetic dipole moments.

The differential cross section for $\tau$ pair production is:

\begin{equation}
\frac{d \sigma}{d  \Omega_{\tau^-}}=
\frac{d \sigma^{0}}{d  \Omega_{\tau^-}}
+\frac{d \sigma^{S}}{d  \Omega_{\tau^-}}+\frac{d \sigma^{SS}}
{d  \Omega_{\tau^-}}+\ldots 
\label{cross1}
\end{equation}

The first three terms come from leading order standard model and effective operator 
(EDM) contributions.
The dots take account for  higher order terms in
the effective Lagrangian that are beyond experimental sensitivity and which are not 
considered in this paper.

The first term of Eq. (\ref{cross1}) represents the spin independent 
differential cross section
\begin{equation}
\frac{d \sigma^0}{d  \Omega_{\tau^-}}=
\frac{\alpha^2 }{16 \ s}\, \beta\, (2-\beta^2 \sin^2 \theta_{\tau^-})
\label{cross0}\end{equation}
where $\alpha$ is the fine structure constant, the squared center of mass energy 
 $s=q^2$ is also the square of the 4-momentum carried by the photon, $\theta_{\tau^-}$ is 
the angle defined by the electron and the $\tau^-$ directions,
 and 
$\gamma=\frac{\sqrt{s}}{2 m_\tau}$,
$\beta=\sqrt{1-\frac{1}{\gamma^2}}$, 
are the dilation factor and $\tau$ velocity, respectively.

The second term 
 $\displaystyle \frac{d\sigma^{S}}{d\Omega_{\tau^-}}$,
involves linear
terms in spin and has no contribution to $CP$-odd observables in our treatment.

The last term of Eq.(\ref{cross1})
is proportional to the product of the spins of both $\tau$'s
and it is written as:

\begin{eqnarray}
\hspace*{-0.6cm}\frac{d\sigma^{SS}}{d \Omega_{\tau^-}}  &=&  \frac{\alpha^2}{16 s} \beta 
 \left( s_+^x s_-^x C_{xx} + s_+^y s_-^y C_{yy} +
s_+^z s_-^z C_{zz} + \right. \nonumber \\
& & \left. (s_+^x s_-^y + s_+^y s_-^x) C_{xy}^+ +
(s_+^x s_-^z + s_+^z s_-^x) C_{xz}^+ + 
(s_+^y s_-^z +s_+^z s_-^y) C_{yz}^+ + \right. \nonumber\\ & &
\left. (\overrightarrow{s}_+ \times \overrightarrow{s}_-)_x C_{yz}^- + 
(\overrightarrow{s}_+ \times \overrightarrow{s}_-)_y C_{xz}^- +
(\overrightarrow{s}_+ \times \overrightarrow{s}_-)_z C_{xy}^-
\right)
\label{csection}\end{eqnarray}

where

\beq\begin{array}{lcl}
C_{xx} =
(2-\beta^2) \sin^2\theta_{\tau^-}
&\hspace*{1cm}&C_{yy} = -\beta^2\, \sin^2\theta_{\tau^-}\\
C_{zz} = (\beta^2+(2-\beta^2) \cos^2\theta_{\tau^-})
&&C_{xy}^- = 2 \beta\left( \sin^2\theta_{\tau^-}\right) \, 
\displaystyle{
\frac{2 m_\tau}{e}}\,d^\gamma_\tau\\
C_{yz}^- = -\gamma\beta  \left(\sin (2\theta_{\tau^-})\right)\,  \displaystyle{
\frac{2 m_\tau}{e}
}
\, d^\gamma_\tau
&&
C_{xz}^+ = \frac{1}{\gamma} \sin (2 \theta_{\tau^-})
\label{cross3}
\end{array}
\eeq

and $C_{xy}^+ = C_{yz}^+ = C_{xz}^- = 0$. $C_{xy}^+$ and $C_{yz}^+$ 
 correlation terms are zero in our hypothesis.
 They are $CP$-even and $P$-odd 
but we only consider photon exchange in the s-channel and there is no
source of $P$ violation to produce these $CP$-even terms. 
$C_{xz}^-$ is zero because it is $CP$-odd and $P$-even,
while the EDM in interference with photon exchange would be 
 $CP$-odd but $P$-odd instead and cannot give contribution to this
term.

The above equations show that the EDM modifies the spin properties of the
produced taus and this translates into the angular distribution of both tau decay
products. As can be seen from Eq.(\ref{cross3}), 
the EDM is the leading contribution to the Normal-Transverse ($y-x$) and
Normal-Longitudinal ($y-z$) correlations. 

Angular asymmetries of the tau decay product distributions  allow 
to select observable EDM effects. In order to enlarge the sensitivity to these terms 
we will sum in all kinematic variables when possible.

The complete  cross section for the process $e^+e^- \rightarrow \gamma 
\rightarrow \tau^+\tau^- \rightarrow h^+\bar{\nu}_\tau h^-\nu_\tau$ can be 
written as a function of the kinematical variables of the hadrons into which each tau 
decays \cite{tsai} as:

\begin{eqnarray}
&&\hspace*{-1cm}d\sigma \left(e^+e^-\rightarrow \gamma
\rightarrow \tau^+\tau^- \rightarrow h^+\bar{\nu}_\tau h^-\nu_\tau\right)=
4\, d\sigma 
\left(e^+e^- \rightarrow \tau^+(\overrightarrow{n}_+^*)
\, \tau^-(\overrightarrow{n}_-^*)\right)\nonumber \\
&&\times \, Br(\tau^+ \rightarrow h^+\bar{\nu}_\tau)
Br(\tau^- \rightarrow h^-\nu_\tau)
\frac{d\Omega_{h^+}}{4\pi}\, \frac{d\Omega_{h^-}}{4\pi} 
\label{eq:cros1}\end{eqnarray}

with
\beq
\overrightarrow{n}_\pm^*= \mp\alpha_\pm \frac{\overrightarrow{q}^{  *}_\pm}{
\arrowvert\overrightarrow{q}^{  *}_\pm\arrowvert} = 
\mp\alpha_\pm(\sin\theta_{\pm}^*\, \cos\phi_\pm,
\sin\theta_{\pm}^*\, \sin\phi_\pm,\cos\theta_{\pm}^*)\nonumber\\
\eeq
$\alpha_\pm$ are 
the polarization parameters of the $\tau$ decay and 
$\overrightarrow{q}_\pm^{ *}$ are the 
momenta of the hadrons with moduli  fixed to
$P_\pm= \displaystyle\frac{m_\tau^2-m_{h^\pm}^2}{2 m_\tau}$. The $*$ means 
that all affected quantities are given in the respective $\tau$-at-rest 
reference frame. Notice that  the boost on the $taus$ is along the $z$  
axis, so the $\phi^*_\pm$ angles do no change when referred 
to the LAB or  the CM 
reference frame and we can just use $\phi_\pm$ to refer to them.  
Both hadron energies
are fixed by energy conservation and the neutrino, in each channel, 
was integrated out in the cross section (\ref{eq:cros1}).
\begin{figure}[hbtp]
\begin{center}
\epsfig{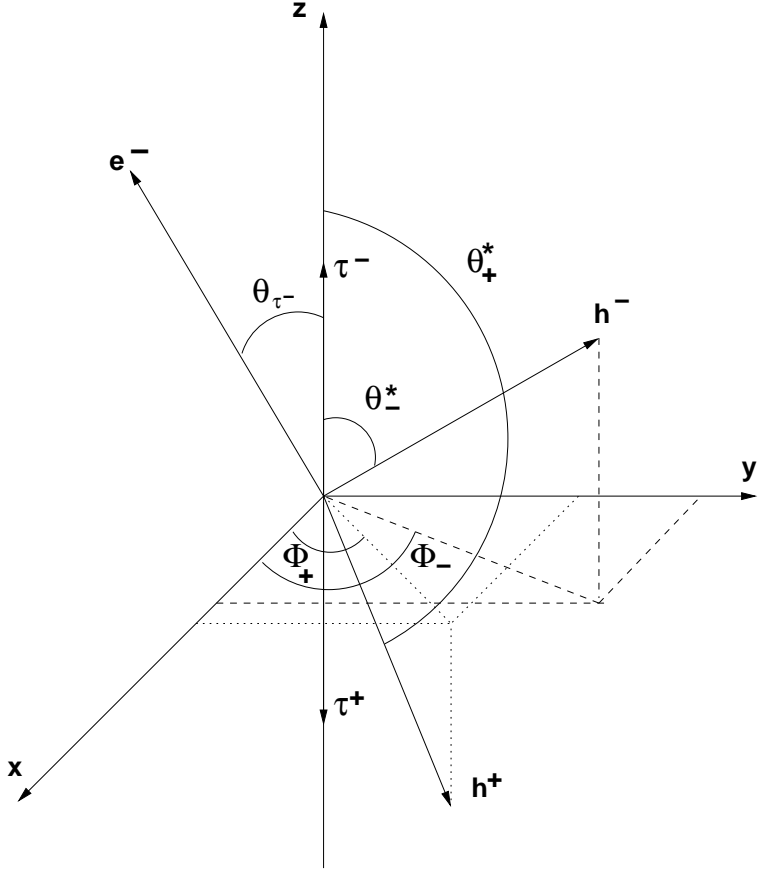}
\end{center}
\caption{Coordinate system for $h^\pm$ production from the $\tau^\pm$}
\label{fig:fig1}
\end{figure}

\subsection{Normal-Transverse correlation azimuthal asymmetry}

We now show how to get an observable proportional to the EDM term 
from the NT correlation.
The correlation terms in the cross section
 depend on  several kinematic variables that we have to take
into account: the CM polar angle $\theta_{\tau^-} $ of production 
of the $\tau^-$ with respect to the
electron, the azimuthal $\phi_{\pm}$ and polar 
$\theta^*_{\pm}$ angles of 
the produced hadrons $h^\pm$ in the $\tau^\pm$
 rest frame (see Fig.\ref{fig:fig1}).
These angles appear in a different way on each term. The $\theta_{\tau^-}$ 
angle enters in  the cross section (spin independent, linear and correlation terms) 
as coefficients 
(such as $C_{xy}^-$, for example) while  the hadron's angles appear in the cross 
section through the $\overrightarrow{n}^*$ vectors. The whole angular dependence of 
each correlation term  is unique and it is this dependence that allows
 to select one of the correlation terms in the cross section.
Indeed, it is by a combination of an integration on the hadronic angles plus, eventually, 
an integration on the $\theta_{\tau^-}$ angle, that one can select 
a polarization or correlation term,
and there, the contribution of the EDM.

For the NT term, for example, this works as follows. 
The integration over the $\tau^-$ variables $d\Omega_{\tau^-}$ erases all 
the information on the EDM in the
Normal-Longitudinal  ($C_{zy}^-$) correlation, together with the $C_{xz}^+$ term of 
the cross section. Then, the cross
section can be written only in terms of  
the surviving correlation terms as:

\begin{eqnarray}
d^4\sigma^{SS}&=&
\frac{\pi\alpha^2\beta}{2\, s}
\, Br(\tau^+ \rightarrow h^+\bar{\nu}_\tau)
Br(\tau^- \rightarrow h^-\nu_\tau) \, \frac{d\Omega_{h^+}}{4\pi}\, 
\frac{d\Omega_{h^-}}{4\pi}\, \times \nonumber\\
&&\Bigg\{\frac{4}{3}(2-\beta^2)\, \left[(n_-^*)_x(n_+^*)_x\right] -
\frac{4}{3}\beta^2\, \left[(n_-^*)_y(n_+^*)_y\right] +\label{eqxy1}\\
&&2\left(\beta^2+(2-\beta^2)\frac{1}{3}\right)\,
\left[(n_-^*)_z(n_+^*)_z\right]+\label{eqxy2}\\
&&\frac{4}{3}\,2\beta\,
 \left[(n_+^*)_x(n_-^*)_y - 
(n_+^*)_y(n_-^*)_x\right]\, \frac{2 m_\tau}{e}\, d^\gamma_\tau\Bigg\}
\end{eqnarray}

Up to this point, linear polarization  terms may also survive (to this order, in fact, it is only a 
longitudinal term that is studied 
in section 5) but a dedicated integration of the hadronic angles ends up with the NT correlation as the 
only  surviving term.

In order to enhance and select the corresponding NT observable we must 
integrate as much as kinematic variables as possible without erasing the 
signal of the EDM. Keeping only azimuthal angles and integrating all 
other variables
we get:

\bea
\frac{d^2\sigma^{SS}}{d\phi_- d\phi_+} 
&=&  -\frac{\pi\alpha^2\beta}{96 s}\,\left(\alpha_-\, \alpha_+\right)\, 
Br(\tau^+ \rightarrow h^+\bar{\nu}_\tau)
Br(\tau^- \rightarrow h^-\nu_\tau)
  \times  \nonumber\\
 & & \Big[(2-\beta^2)\cos(\phi_-)\; \cos(\phi_+)-\beta^2\, \sin(\phi_-)\; \sin(\phi_+)+ 
 \label{eqxy3} \\
&&2\beta\, \sin (\phi_-^* - \phi_+^*)\, \frac{2
m_\tau}{e}\,d^\gamma_\tau  \Big]
\eea

Now, to get only sensitivity to the EDM in the NT correlation we can define the 
azimuthal asymmetry as:

\begin{equation}
A_{NT} = \frac{\sigma^+_{NT} - \sigma^-_{NT}}{\sigma^+_{NT} + \sigma^-_{NT}}
\end{equation}

where

\begin{equation}
\sigma^\pm_{NT} = 
\displaystyle{
\int_{w \stackrel{_>}{_<} 0}
\frac{d^2\sigma}{d\phi_- d\phi_+}
} \,d\phi_-\,d\phi_+\ , \quad \mbox{with}\quad \quad w = \sin(\phi_--\phi_+)
\end{equation}

so that one gets that the Normal-Transverse correlation azimuthal asymmetry is:
\beq
A_{NT} =- \alpha_-\alpha_+ \, \frac{\pi \beta}{4(3 - \beta^2)} \, 
\frac{2 m_\tau}{e}\, d^\gamma_\tau
\label{antphi+-}\eeq

It is easy to verify that all other terms in the cross section, i.e. the 
spin independent ones ($\displaystyle \sigma^0$), the ones coming with 
the linear  polarization ($\sigma^{S}$) and those not relevant of the
spin-spin correlation term  (Eq.(\ref{eqxy3}) in $\sigma^{SS}$) are
eliminated when we integrate in the way we have shown above. Notice that this
integration procedure also erases any possible contribution coming from 
the $CP$-even term $C_{xy}^+$ (zero in our  approach) of the NT 
polarization.
This means that the only source for this azimuthal asymmetry is 
exactly the term $C_{xy}^-$ we are interested in, so that we
have ended up with a genuine $CP$-odd Normal-Transverse correlation 
observable which is directly proportional to the EDM. 

\subsection{Normal-Longitudinal correlation asymmetry}

In a similar way, we can now define an observable related 
to the Normal-Longitudinal correlation term. In this case the angular 
dependence on the decay product of both $\tau$ is different:

\begin{eqnarray}
\hspace*{-.5cm}\left.\frac{d\sigma}{d\left(\cos\theta_{\tau^-}\right)}
\right|_{C^-_{yz}}&=&
\frac{\pi\alpha^2\beta^2\gamma}{2\, s} \frac{2m_\tau}{e}\; d^\gamma_\tau\; 
\sin(2\theta_{\tau^-})\,
\left[(n^*_-)_y(n^*_+)_z-(n^*_-)_z(n^*_+)_y\right]\times \nonumber \\
&& Br(\tau^+ \rightarrow h^+\bar{\nu}_\tau)
Br(\tau^- \rightarrow h^-\nu_\tau)\, \frac{d\Omega_{h^+}}{4\pi}\, 
\frac{d\Omega_{h^-}}{4\pi}\label{nl1}
\end{eqnarray}

We can not just integrate over the $\theta_{\tau^-}$ variable because this
 erases all the information on
the EDM of Eq.(\ref{nl1}), so we must do a forward-backward (with respect to the
$e^-$ direction) integration of the $\tau¯$ :
\beq
d\sigma(\pm)\equiv \left[\; \int_{0}^1d(\cos\theta_{\tau^-})\pm
\int_{-1}^0d(\cos\theta_{\tau^-})\, \right]d\sigma
\eeq

Then, from Eq.(\ref{csection}), 
only terms on $\sin(2\theta_{\tau^-})$ survive for $d\sigma(-)$,
\bea
d\sigma(-)^{SS}&\equiv&
\left[\; \int_{0}^1d(\cos\theta_{\tau^-})-
\int_{-1}^0d(\cos\theta_{\tau^-})\, \right]d\sigma^{SS}
=\frac{2\pi \alpha^2 \beta}{3\, s}\times \nonumber \\
&&\left\{\left[(n^*_-)_y(n^*_+)_z-(n^*_-)_z(n^*_+)_y\right]\, \gamma\beta\, 
\frac{2m_\tau}{e}\;  d^\gamma_\tau \right. \\
&&\left.+\left[(n^*_-)_x(n^*_+)_z+(n^*_-)_z(n^*_+)_x\right]\frac{1}{\gamma}\right\}\times
\label{irrelevante}\\
&&Br(\tau^+ \rightarrow h^+\bar{\nu}_\tau)
Br(\tau^- \rightarrow h^-\nu_\tau) \, \frac{d\Omega_{h^+}}{4\pi}\, 
\frac{d\Omega_{h^-}}{4\pi}\label{nl2}
\eea

Again, this is not enough to select the NL correlation. 
The dependence on the $n^*$  produces an angular dependence on the angles 
of the form 
\bea
(n^*_-)_y(n^*_+)_z&-&(n^*_-)_z(n^*_+)_y=\nonumber \\
&&\alpha_+\alpha_-\left(\sin\theta^*_+\,\cos\theta^*_-
\, \sin\phi_+ - \cos\theta^*_+\, \sin\theta^*_-\, \sin\phi_-
\right)\label{menos}\\
(n^*_-)_x(n^*_+)_z&+&(n^*_-)_z(n^*_+)_x=\nonumber \\
&&-\alpha_+\alpha_-\left(\sin\theta^*_-\,\cos\theta^*_+
\, \cos\phi_- + \cos\theta^*_-\, \sin\theta^*_+\, \cos\phi_+\right)
\eea

By an appropriate integration of the remaining variables 
one may  get rid of the
irrelevant (for our purposes) term Eq.(\ref{irrelevante}) and get sensitivity to 
the EDM. For example one may integrate in such a way as to 
select the first term in Eq.(\ref{menos}). This can be done computing

\begin{equation}
\hspace*{-.5cm}
\sigma_{NL}(-)^{\pm}_- = 
\displaystyle{
\int_{w \stackrel{_>}{_<} 0}
\frac{d^2\sigma}{d(\cos\theta^*_-) d\phi_+}
} \,d(\cos\theta^*_-)\,d\phi_+\ ,\hspace*{.25cm} \mbox{with}\quad w = \cos\theta^*_- \sin\phi_+
\end{equation}
which amounts  to calculate

\bea 
&&\hspace*{-.9cm}\sigma_{NL}(-)^+_\mp\equiv\left[\,
\left(\int_0^\pi\d\phi_\pm\int_0^1d(\cos\theta^*_\mp)+
\int_\pi^{2\pi}d\phi_\pm\int_{-1}^0d(\cos\theta_\mp)\right)\,
\right]
\frac{d^2\sigma^{SS}}{d(\cos\theta^*_\mp)d\phi_\pm}\\
&&\hspace*{-.9cm}\sigma_{NL}(-)^-_\mp\equiv\left[\,
\left(\int_0^\pi\d\phi_\pm\int_{-1}^0d(\cos\theta^*_\mp)+
\int_\pi^{2\pi}d\phi_\pm\int_{0}^1d(\cos\theta_\mp)\right)\,
\right]
\frac{d^2\sigma^{SS}}{d(\cos\theta^*_\mp)d\phi_\pm}
\eea

so that 

\bea
&&\sigma_{NL}(-)_-^+ - \sigma_{NL}(-)_-^-=\nonumber \\
&&\frac{\pi\alpha^2\beta^2\gamma}{6 s}(\alpha_+\alpha_-)\, 
\frac{2m_\tau}{e}\;  d^\gamma_\tau\; Br(\tau^+ \rightarrow h^+\bar{\nu}_\tau)
Br(\tau^- \rightarrow h^-\nu_\tau)
\eea
Then one can construct the corresponding asymmetry as:

\begin{equation}
A^+_{NL} = \frac{\sigma_{NL}(-)_+^+ - \sigma_{NL}(-)_+^-}{\sigma_{NL}(-)_+^+ +\sigma_{NL}(-)_+^-}=
\frac{\beta\gamma}{4(3-\beta^2)}\, 
\alpha_+\alpha_-\, \frac{2m_\tau}{e}\, d^\gamma_\tau
\label{anl}
\end{equation}

Notice that a similar asymmetry can be build by interchanging 
$\phi_+\leftrightarrow\phi_-$, $\theta^*_-\leftrightarrow\theta^*_+$ 
 and the $\pm$ signs of the $\sigma$ sub-indexes 
in the above expressions:

\begin{equation}
A^-_{NL} = \frac{\sigma_{NL}(-)_-^+ - 
\sigma_{NL}(-)_-^-}{\sigma_{NL}(+)_-^+ +\sigma_{NL}(+)_-^-}=-A_{NL}^+
\label{anl2}
\end{equation}

As $C_{yz}^+ = 0$ in our approach, along this process of integration only the $CP$-odd $C_{yz}^-$ term 
survives. We have verified that all other terms of Eq. (\ref{cross3}) are 
annihilated in the definition of this asymmetry. 

Differently to what happened in
the NT asymmetry  of Eq. (\ref{antphi+-}), the NL asymmetry defined in Eq.
(\ref{anl}) is not a genuine $CP$-violation observable because it can get 
possible 
contributions from the $CP$-even sector $C_{yz}^+$ (zero in our case) 
of the cross section. To get a genuine $CP$-odd observable one has to test 
in this 
case  both $\tau$'s decaying into the same kind of hadrons 
($\alpha_-=\alpha_+\equiv\alpha_h$) and then define the NL asymmetry,

\beq
A_{NL}=\frac{1}{2}\left(A^+_{NL}-A^-_{NL}\right)=
\frac{\beta\gamma}{4(3-\beta^2)}\, \alpha_h^2\, \frac{2m_\tau}{e}\, d^\gamma_\tau
\eeq

that exactly  tests the $CP$-odd sector  ({\it i.e.} $C_{yz}^-$ only) of the 
Normal-Longitudinal correlation.


Notice that the above expressions for the asymmetries are  linear in the 
$\alpha_\pm$ factors so that 
we can consider all the decaying channels of both  $\tau$'s (to $\pi$, 
$\rho$...) in order 
to increase statistics and enlarge the signal.

\section{Observables at the $\Upsilon$ resonances}

All these ideas can be applied for $e^+e^-$ collisions at the $\Upsilon$ peak 
where the $\tau$ pair production is  mediated by the resonance: 
$e^+e^- \rightarrow \Upsilon \rightarrow \tau^-\tau^-$.
At the $\Upsilon$ production energies  we 
 have an important  tau pair production rate.
We are interested in $\tau$ pairs produced by the 
 decays of the $\Upsilon$ resonances, therefore 
we can use $\Upsilon(1S)$, $\Upsilon(2S)$ and $\Upsilon(3S)$ where the decay rates
into tau pairs have been measured. At the  
$\Upsilon(4S)$ peak, although  it decays dominantly into $B\overline{B}$, high luminosity
B-Factories have an important direct tau pair production. Except for this last case,
that can be studied with the results of the preceding sections,  we assume that
the resonant diagrams (b) and (d) of Fig. 1. dominate the process 
on the $\Upsilon $ peaks.
 The Breit-Wigner propagator of the $\Upsilon$ is

\beq
P_\Upsilon(s) = \frac{1}{(s-M_\Upsilon^2) + i M_\Upsilon \Gamma_\Upsilon}
\eeq

The $F_\Upsilon (q^2)$  vector form factor is defined as

\beq
\langle \Upsilon(w,\mbox{\boldmath $q$}) | 
\bar{\psi}_b\gamma_\mu\psi_b(0)|0\rangle 
= F_\Upsilon (q^2) \epsilon^*_\mu(w,\mbox{\boldmath $q$})
\eeq

with $\epsilon^*_\mu(w,\mbox{\boldmath $q$})$ the polarization 
four-vector. This form factor can be related to the partial width of $\Upsilon 
 \rightarrow e^+e^-$,

\beq
\Gamma_{ee} = \frac{1}{6\pi}Q_b^2
\frac{(4\pi\alpha)^2}{M_\Upsilon^4} |F_\Upsilon(M_\Upsilon)|^2 \frac{M_\Upsilon}{2}\; ,
\eeq
where $Q_b = - 1/3$ is the electric charge of the $b$ quark.
Notice that all the hadronic uncertainties in our process are 
included in this unique form factor.

With this parameterization, the amplitudes $A_b$ and $A_d$  for the tau pair 
production diagrams (see Fig. 1.)
 through the $\Upsilon$ can be related to those of the direct 
production ($A_a$ and $A_c$) as:
\beq
A_{b\choose d}=A_{a\choose c}\cdot H(s)\; ,\quad \mbox{with }\quad H(s)\equiv
\frac{4\pi \alpha Q_b^2}{s}\, |F_\Upsilon(s)|^2\, P_\Upsilon(s)\, ,\label{amplitud}
\eeq
so that the tau pair production at the $\Upsilon$ peak introduces the same tau 
polarization matrix terms as the direct production with $\gamma$ exchange 
(diagrams (a) and (c)). The only difference is the overall factor $|H(s)|^2$
introduced in the cross section which is responsible for the enhancement 
at the resonant energies,
\beq 
H(M_\Upsilon^2)=\frac{4\pi \alpha Q_b^2}{M_\Upsilon^2}
\frac{|F_\Upsilon(M_\Upsilon^2)|^2}{i\, \Gamma_\Upsilon\, M_\Upsilon}=-i \, 
\frac{3}{\alpha} Br\left(\Upsilon \rightarrow e^+e^-\right) \label{factor}
\eeq

>From Eqs. (\ref{amplitud}) and (\ref{factor}) it is easy to show that, at the
$Upsilon$ peak,  the 
interference of diagrams (a) and (d) plus the interference of diagrams (b) 
and (c) is exactly zero and so it is the interference of diagrams (a) and (b).
Finally, the only contributions with  EDM in polarization terms  
come  with the interference of diagrams (b) and (d), while 
diagram (b) squared gives the leading contribution to the cross section.

 The computations we did following  Eqs.(\ref{cross1}), (\ref{cross0}) and 
(\ref{csection}) can be repeated here, and finally we obtain no changes in the 
asymmetries:
 the only difference  is in the value of the resonant production cross section 
at the $\Upsilon$ peak that is multiplied by the overall factor
$|H(M_\Upsilon^2)|^2$ given in Eq. (\ref{factor}).
In this way all the asymmetries defined by Eqs. (\ref{antphi+-}) 
and (\ref{anl}) 
do not change, and their expressions  at the 
$\Upsilon$ peak are the same as before.

In fact,  one can take the four diagrams (a,b,c,d) together and still get 
the same results of this section and section 3. Energies {\it off}
or {\it on} the resonance will automatically select the significant diagrams.

\section{Imaginary EDM observables}

 
The imaginary part of the EDM does not appear in the effective 
Lagrangian approach and deserves a separate treatment.
This is a $T$-even quantity and it 
can contribute to the cross section  in the $CP$-odd components of the transverse (within the 
production plane) 
 $(s_+-s_-)_T$ and longitudinal $(s_+-s_-)_L$ polarizations.
 These are $P$-odd observables so that the interference of the EDM
 with photon exchange will originate a non vanishing contribution. As a consequence,
  the leading contribution to
 the single polarization terms in the cross section are given by:

\beq
\frac{d\sigma^{S}}{d\Omega_{\tau^-}}=\frac{\alpha^2 \beta^2}{16\, s} 
\left[
(s_-^x-s_+^x)C_x+(s_-^z-s_+^z)C_z\right]
\label{cross2}
\eeq

where
\beq 
C_x=-\gamma\sin(2\theta_{\tau^-})\, \frac{2m_\tau}{e}\, \im{d_\tau^\gamma},
\quad C_z=2\, \left(\sin^2\theta_{\tau^-}\right)\, \frac{2m_\tau}{e}\; \im{d_\tau^\gamma}
\eeq

Contributions to the $d\sigma^S$ could also come from the $CP$-odd interference of the 
real part of the EDM with absorptive parts and
from the $CP$-even $\gamma-Z$ 
interference.

>From Eq. (\ref{cross2}), one can see that the transverse polarization
term has an angular dependence of the form 
\beq
(\alpha_+\sin\theta^*_+\cos\phi_+
+ \alpha_-\sin\theta_-^*\cos\phi_-)\, \sin \left(2\theta_{\tau^-}\right) 
\eeq
Integrating the cross section in all angles 
except $\theta_{\tau^-}$ and $\phi_\pm$ we can define an asymmetry  

\begin{equation}
A_{T}^\pm = \frac{
\sigma(+)^\pm - \sigma(-)^\pm
}{\sigma(+)^\pm + \sigma(-)^\pm}
\end{equation}

where

\begin{eqnarray}
\sigma(+)^\pm &\equiv& 
\int_{w>0}
\frac{d^2\sigma}{d(\cos\theta_{\tau^-}) d\phi_\pm}
 \,d(\cos\theta_{\tau^-})\, d\phi_\pm,\nonumber\\ 
\sigma(-)^\pm &\equiv& 
\int_{w<0}
\frac{d^2\sigma}{d(\cos\theta_{\tau^-}) d\phi_\pm} \, d(\cos\theta_{\tau^-})\,d\phi_\pm
\end{eqnarray}
and $w = \sin \left(2\theta_{\tau^-}\right)\cos \phi_\pm$, so that

\beq
A^\pm_T = -\frac{\beta\gamma}{2(3 - \beta^2)}( \alpha_\pm) \, \frac{2 m_\tau}{e}\, 
\im{d^\gamma_\tau}
\label{at}
\eeq

This asymmetry receives also standard contributions form the $\gamma-Z$
interference term to $(s_-^x+s_+^x)$. We  want to isolate  the EDM 
signal only, so one  has to define a genuine $CP$-odd transverse asymmetry. 
We assume that both
$\tau$'s decay into the same kind of hadrons 
($\alpha_-=\alpha_+\equiv\alpha_h$),
\beq
A_{T}=\frac{1}{2}\left(A^+_{T}+A^-_{T}\right)=
-\frac{\beta\gamma}{2(3 - \beta^2)}( \alpha_h) \, \frac{2 m_\tau}{e}\, 
\im{d^\gamma_\tau}
\label{at+-}
\eeq
which eliminates the standard model contribution.

A similar procedure can be done for the longitudinal polarization. In that case 
the angular dependence is of the form:
\beq \left(\alpha_-\cos\theta^*_-+\alpha_+\cos\theta^*_+\right)\,
\sin^2\theta_{\tau^-}
\eeq
so that $\theta_{\tau^-}$ and $\phi_\pm$ variables can be integrated out without
erasing the signal. The asymmetry is then defined to be
\beq
A_{L}^\pm=\frac{\sigma_L(+)^\pm-\sigma_L(-)^\pm}{\sigma_L(+)^\pm+\sigma_L(-)^\pm}=
\frac{\beta}{3-\beta^2}\, (\alpha_\pm)\, \frac{2m_\tau}{e}\,
\im{d^\gamma_\tau}\label{alpm}
\eeq
with 
\beq 
\hspace*{-.5cm}\sigma_L(+)^\pm\equiv \int_0^1 d(\cos\theta^*_\pm)\,
\frac{d\sigma}{d(\cos\theta^*_\pm)},\qquad 
\sigma_L(-)^\pm\equiv \int_{-1}^0 d(\cos\theta^*_\pm)\,
\frac{d\sigma}{d(\cos\theta^*_\pm)}
\label{al}
\eeq

>From these observables one can again define a genuine 
$CP$-odd longitudinal asymmetry 
\beq
A_L=\frac{1}{2}\left(A_L^++A_L^-\right)=
\frac{\beta}{3-\beta^2}\, (\alpha_h)\, \frac{2m_\tau}{e}\, \im{d^\gamma_\tau}
\label{al+-}
\eeq
that erases standard model contributions $(s_-^z+s_+^z)$ coming 
from $\gamma-Z$ interference.

In each one of these cases we have verified that
 all the other terms in the cross section do not contribute
to the asymmetries and are eliminated 
when we integrate in the  angles.
When measuring these single-tau asymmetries, for each decaying 
channel of the observed $tau$, one may increase the statistics by summing up 
over the $\pi$, 
$\rho$... semileptonic decay channels of the $\tau$ for which the angular 
distribution is not observed.

Let us finally point out that the $CP$-odd Transverse and Longitudinal 
polarization asymmetries
of Eqs. (\ref{at+-}) and  (\ref{al+-})
 get a contribution from the EDM real part 
through its interference with the $Z$-exchange.
They 
give, however, a vanishing small contribution. These terms are
\bea
\hspace*{-.5cm}^{(\gamma-Z)}A_T^\pm&=&
\frac{\beta\gamma}{2(3 - \beta^2)}( \alpha_\pm) \, \frac{2 m_\tau}{e}\, 
\overbrace{\left[\frac{s\Gamma_Z M_Z}{(s-M_Z^2)^2+(\Gamma_ZM_Z)^2}
\frac{v_e\, v_\tau}{4s_w^2\,
c_w^2}\right]}^{R} \re{d^\gamma_\tau}\\
\hspace*{-.5cm}^{(\gamma-Z)}A_L^\pm&=&
-\frac{2\beta}{3-\beta^2}\, (\alpha_\pm)\, \frac{2m_\tau}{e}\,
\underbrace{\left[\frac{s\Gamma_Z M_Z}{(s-M_Z^2)^2+(\Gamma_ZM_Z)^2}\frac{v_e\, v_\tau}{4s_w^2\,
c_w^2}\right]}_{R} \re{d^\gamma_\tau}
\eea
and they are suppressed, respect to the single
 photon exchange (\ref{at}) and (\ref{alpm}),
 by the factor $R$ of the order 
$(\frac{q^2}{M_Z^2}  \frac{\Gamma_Z}{M_Z})$. At Upsilon energies this
factor is of the 
order $6\cdot 10^{-7}$, so that one can safely conclude that ($\gamma-Z$) contributions can be 
neglected within the present experimental accuracy.

\section{Bounds on the EDM and final remarks }
We can now estimate the bounds on the EDM that can be achieved using these 
observables. We assume a conservative set of data of $10^6$ ($10^7$) tau pairs produced
from all Upsilon resonances. This would presume a collection of $4\times 10^7$
($4\times 10^8$) $\Upsilon(1S)$ for example. The bounds one gets for the 
EDM are:
\bea
&&\hspace*{-.75cm}\mbox{NT asymmetry and }\pi^\pm \mbox{ $tau$ decay channel: }
\left|d^\gamma_\tau\right|<\left\{\begin{array}{l}
1.5\times 10^{-16}\ {\it e \, cm}\\
(4.9\times 10^{-17}\ {\it e \,cm})\end{array}\right.\\
&&\hspace*{-.75cm}\mbox{NL asymmetry and }\pi^\pm \mbox{ $tau$ decay channel: }
\left|d^\gamma_\tau\right|<\left\{\begin{array}{l}
1.7\times 10^{-16}\ {\it e \, cm}\\
(5.4\times 10^{-17}\ {\it e \, cm})\end{array}\right.
\eea

While for the imaginary part of the EDM the bounds are:
\bea
&&\hspace*{-.75cm}\left.\begin{array}{l}\mbox{T asymmetry and }\\
\pi^\pm \mbox{ $tau$ decay channels }\end{array}\right\}:\quad 
\left|\im{d^\gamma_\tau}\right|<\left\{\begin{array}{l}
8.3\times 10^{-17}\ {\it e \, cm}\\
(2.6\times 10^{-17}\ {\it  e \, cm})\end{array}\right.\\
&&\hspace*{-.75cm}\left.\begin{array}{l}\mbox{L asymmetry and }\\
\pi^\pm \mbox{ $tau$ decay channels }\end{array}\right\}:\quad
\left|\im{d^\gamma_\tau}\right|<\left\{\begin{array}{l}
1.2\times 10^{-16}\ {\it e \, cm}\\
(3.7\times 10^{-17}\ {\it  e\, cm})\end{array}\right.
\eea

To  conclude, we would like to point out:

\begin{itemize}

\item[-] with low energy data  we may have  an independent analysis 
of the EDM from that obtained with high energy data,

\item[-] low energy data makes possible a clear separation of the effects
coming from the electromagnetic-EDM and the weak-EDM,

\item[-] high statistics can  compensate the suppression factor
$q^2/\Lambda^2$ in the low energy regime for the effective operators,

\item[-] low energy observables, as defined in this paper, are different and complementary
to the high energy ones studied in the past, and 

\item[-] competitive bounds can be obtained  from this new set of low energy observables.
\end{itemize}
\begin{ack} 
This work has been supported by CONICYT-6052-Uruguay, by
MCyT Spain, under
the grants BFM2002-00568 and FPA2002-00612, and by the 
Agencia Espa\~nola de Cooperaci\'on Internacional. 
\end{ack} 


\begin{thebibliography}{20} 
 
\bibitem{pdg} S.~Eidelman {\it et al.}, \pl{B592}{2004}{1}; K.~Inami {\it et al.} [BELLE Collaboration], 
\pl{B551}{2003}{16}. 

\bibitem{nt} J.~Bernab\'eu, G.A.~Gonz\'alez-Sprinberg and J.~Vidal, 
\pl{B326}{1994}{168}.

\bibitem{heidel} W.~Bernreuther, U.~Low, J.~P.~Ma and O.~Nachtmann,
Z.\ Phys.\ C {\bf 43}, 117 (1989).

\bibitem{l3} M. Acciarri {\it et al.} [L3 Collaboration], \pl{B434}{1998}{169}. 

\bibitem{ao} D.~Buskulic {\it et al.}  [ALEPH Collaboration],
Phys.\ Lett.\ B {\bf 346}, 371 (1995);
K.~Ackerstaff {\it et al.}  [OPAL Collaboration],
Z.\ Phys.\ C {\bf 74}, 403 (1997); H.Albrecht {\it et al.} [ARGUS Collaboration], \pl{B485}{2000}{37}.

\bibitem{paco} F. del Aguila and M. Sher, \pl{B252}{1990}{116}.

\bibitem{grif} J.A. Grifols and A. Mendez, \pl{B255}{1991}{611} and Erratum 
\pl{B259}{1991}{512}.

\bibitem{masso} R. Escribano and E. Masso, \pl{395}{1997}{369}.

\bibitem{nu}
Y.~Fukuda {\it et al.} [Super-Kamiokande Collaboration],
Phys.\ Rev.\ Lett.\ {\bf 81} (1998) 1562. 

\bibitem{buch}W.~Buchmuller and D.~Wyler, \np{B268}{1986}{621}; 
C.N.~Leung, S.T.~Love and S.~Rao, \zp{C31}{1986}{433}; %
M. Bilenky and A. Santamaria, \np{B420}{1994}{47}. 

\bibitem{arcadi}  G.A.~Gonz\'alez-Sprinberg, A.~Santamaria, J.~Vidal, \np{B582}{2000}{3}.

\bibitem{500}
W.~Bernreuther, O.~Nachtmann and P.~Overmann,
Phys.\ Rev.\ D {\bf 48} (1993) 78.

\bibitem{nuria}J. Bernab\'eu and N. Rius,\pl{B232}{1989}{127};
J. Bernab\'eu, N. Rius and A. Pich,\pl{B257}{1991}{219}.

\bibitem{nos} J.~Bernab\'eu, G.A.~Gonz\'alez-Sprinberg, M.~Tung and J.~Vidal, 
\np{B436}{1995}{474}.

\bibitem{kuhn}
J.~H.~Kuhn, Phys.\ Lett.\ B {\bf 313} (1993) 458.


\bibitem{tsai} Y.S.~Tsai \pr{D4}{1971}{2821}.

\end{thebibliography}
\end{document}